\documentclass[aps,pra,twocolumn,groupedaddress,10pt]{revtex4}
\usepackage{amsmath}
\usepackage{amssymb}
\usepackage{graphics}
\usepackage{graphicx}
\usepackage{pstricks}
\usepackage{bbold}
\providecommand{\av}[1]{\left\langle #1\right\rangle}

\providecommand{\ket}[1]{|#1\rangle}
\providecommand{\bra}[1]{\langle#1|}
\providecommand{\kebra}[2]{\ket{#1}\bra{#2}}

\providecommand{\tr}[1]{\textrm{tr}\left\{#1\right\}}

\providecommand{\colvec}[2]{\left(\begin{array}{c}
\mathbf{#1}\\
\mathbf{#2}
\end{array}\right)}

\providecommand{\eq}[1]{Eq.~(\ref{#1})}
\newcommand{\ie}{{\it{i.e.}}}
\newcommand{\eg}{{\it{e.g.}}}
\newcommand{\etal}{{\it{et al.}}}

\begin{document}
\title{Reconstructing the quantum state of oscillator networks with a single qubit}
\author{Tommaso Tufarelli$^1$,  Alessandro Ferraro$^1$, M. S. Kim$^2$, Sougato Bose$^1$}
\affiliation{$^1$Department of Physics and Astronomy, University College London, Gower Street, London WC1E 6BT, UK\\
$^2$QOLS, Blackett Laboratory, Imperial College London, SW7 2BW, UK}
\begin{abstract}
We introduce a scheme to reconstruct arbitrary states of networks composed of quantum oscillators---e.g., the motional state of trapped ions or the radiation state of coupled cavities. The scheme involves minimal resources and minimal access, in the sense that it {\it i)} requires only the interaction between a one-qubit probe and a single node of the network; {\it ii)} provides the Weyl characteristic function of the network directly from the data, avoiding any tomographic transformation; {\it iii)} involves the tuning of only one coupling parameter. In addition, we show that a number of quantum properties can be extracted without full reconstruction of the state. The scheme can be used for probing quantum simulations of anharmonic many-body systems and quantum computations with continuous variables. Experimental implementation with trapped ions is also discussed and shown to be within reach of current technology.
\end{abstract}
\maketitle
Coupled harmonic and anharmonic oscillators constitute the building blocks of mathematical models that are ubiquitous in physics. Quantum systems are no exception and, in fact, the studies on coupled quantum oscillators trace back to the origin of quantum physics itself. The emergence of quantum information science has added renewed interest in these continuous-variable systems \cite{cv_rev}. For example, the possibility to exert exquisite experimental control over {\it travelling} oscillator modes led to novel applications in quantum optical communication \cite{cv_comm}. However, these experiments involve solely a limited number of modes, whereas both the investigation on many-body models and the quest for advanced quantum information tasks call for the realization of more complex bosonic networks. Interestingly, some alternative experimental settings are now reaching maturity for implementing these networks, thanks to unprecedented ability to manipulate {\it confined} quantum modes \cite{note1}. In fact, trapped ions, cavity QED, circuit QED, or nanomechanical oscillators have been proposed to realize quantum simulators of many-body systems whose properties are beyond reach of purely theoretical and numerical investigations \cite{cv_qs}. 
Suggestions for the use of these platforms for continuous-variable quantum computation \cite{cv_qc1} have also been recently put forward \cite{cv_qc2}. In addition, general physical concepts, related to entropy-area laws \cite{area} or quantum thermodynamics \cite{qtherm}, have been extensively analyzed for oscillator networks and could be amenable for experimental testing.

Despite the aforementioned proposals, fundamental tools that still lack in this context are {\it minimal} and {\it feasible} schemes to reconstruct the quantum state of oscillator networks---a necessary step for probing the validity of quantum simulations and computations. In general, a reconstruction scheme---also dubbed quantum state tomography---tries to estimate a quantum state using measurements on an ensemble of identical copies of it. Considering travelling modes a huge research effort has been made in the past years and quantum tomography is now standard \cite{qtomo}. However, the latter is based on the measure of quadrature signals, which are unavailable for confined quantum modes. To face this obstacle, in the case of a single oscillator, many alternative schemes have then been put forward, relying on interrogating the system either with a discrete-variable (qubit) \cite{ion_rmp,DVProbe} or a continuous-variable \cite{CVProbe} probe. However, the adaptation of these schemes to the relevant case of a network of many oscillators has been vastly overlooked (see Refs.~\cite{ions} for some details). 
In quantum tomography of travelling optical fields, the state reconstruction of a $N$-mode field requires  the ability to perform joint measurements of $N$ arbitrary field quadratures (one for each mode). Similarly, if we wanted to reconstruct the joint state of $N$ oscillators with a probe-mediated method, we might introduce $N$ auxiliary probes. This approach, requiring \textit{maximal access} to the network, can quickly become impractical as the number of oscillators is increased. In fact, accessibility constraints often plague experiments and thus the question of extracting information with only partial access---a non-trivial inverse problem involving an interacting many-body system---assumes also a practical relevance. For example, two recent experiments with trapped ions have probed the dynamics of the simplest possible network, composed of two oscillators \cite{2ions}. There, only one of the oscillators could be probed, imposing partial access to the system. In general, it is thus desirable to design state reconstruction protocols that involve a smaller number of resources, as compared to the straightforward extension of the single-oscillator schemes. We introduce here one such protocol, that solves these major drawbacks by requiring only \textit{minimal access} to the network.
In particular, it involves only the  \textit{interaction between one qubit-probe and one constituent of the network.} In addition, the method provides directly the Weyl characteristic function of the system, avoiding the massive post-processing of noisy data common to many reconstruction schemes---a benefit that considerably eases its implementation. 

\begin{figure}[t!]
	\center{\includegraphics[width=.8\linewidth]{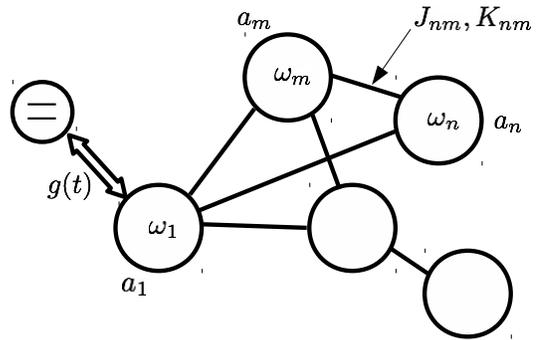}\caption{Graphical representation of the Hamiltonian of \eq{HTOT}. A qubit (two-level system) is tunably coupled [$g(t)$] to a single constituent ($a_1$) of an oscillator network. The constituents of the network ($a_j$, with $j=1,...,N$) are in turn harmonically coupled ($J_{n,m},K_{n,m}$).\label{network}}
	}\end{figure}

We consider a generic oscillator network in an unknown state---possibly being an eigenstate of some simulated anharmonic model, or an intermediate state of a quantum computation. Regardless the previous dynamics, we suppose that, from a certain time $t=0$, the oscillators interact only harmonically. In addition, a single qubit can interact with a single (fixed) oscillator via a tunable bilinear coupling (see Fig.\ref{network}). Such a network-probe system is then let evolve for a certain period of time, allowing part of the information about the network state to be transfered into the qubit. Afterwards, only the qubit is measured. Repeating the procedure it is possible to reconstruct the state of the whole network, by solely tuning the profile of the interaction strength.

The paper is structured as follows. In Sec.~\ref{sec_ham} we introduce the Hamilonian model and solve the time dynamics of the system. We will see in Sec.~\ref{sec_dis} that the system dynamics performs arbitrary qubit-controlled multimode displacements of the network. This, in turn, allows to implement a complete reconstruction of the network state by solely measuring the qubit, as shown in Sec.~\ref{sec_qsr}. The effects of the major sources of noise will be taken into account in Sec.~\ref{sec_noise} (see also Appendices \ref{appendix1} and \ref{appendix2}). We conclude the paper by considering the example of a linear chain of oscillators (Sec.~\ref{linearchain}) and discussing possible implementations of our scheme (Sec.~\ref{sec_exp}).

\section{Hamiltonian and time evolution} \label{sec_ham}
We consider a network-qubit system whose total Hamiltonian at $t\geq0$ is given (in a frame rotating with the free Hamiltonian of the qubit) by
\begin{align}
	&H(t)=H_0+H_\textrm{int}(t),\label{HTOT}\\
&H_0=\sum_{n=1}^N\omega_na^\dagger_na_n+\sum_{n<m}J_{nm}\left(a_na_m^\dagger+a_n^\dagger a_m\right)+\nonumber\\
&\phantom{H_0}+\sum_{n<m}K_{nm}\left(a_na_m+a_n^\dagger a_m^\dagger\right),\label{H_0}\\
&H_\textrm{int}(t)=g(t)\sigma_3\left(a_1+a_1^\dagger\right),\label{time-varying}
\end{align}
where $N$ is the number of oscillators, $a_n$ the bosonic annihilation operator for the $n$-th oscillator, $\omega_n$ the corresponding local frequency, $J_{nm}$ and $K_{nm}$ the interaction strengths between the $n$-th and the $m$-th oscillator, and $g(t)$ the time-varying coupling strength between the qubit and a single oscillator of the network, which we label $n=1$. The operator $\sigma_3$ is a generic Pauli operator for the qubit, belonging to a right-handed tern $\sigma_1,\sigma_2,\sigma_3$, with $[\sigma_i,\sigma_j]=2i\sum_k\epsilon_{ijk}\sigma_k$. 
The mutual interactions between the oscillators and the qubit in Eqs.~(\ref{H_0}) and (\ref{time-varying}) suggest that information can propagate from any node of the network to the first one (and  the qubit), in turn permitting the reconstruction of the whole network state by accessing only the first node. However, these mutual interactions also yield the dispersion of any signal along the network. Thus, before giving the explicit reconstruction method, we first have to solve the inverse problem of unravelling the intricate dynamics of this interacting system. \subsection{Normal modes decomposition}\label{normal-modes}
\begin{figure}
	\begin{center}
		\includegraphics[width=.8\linewidth]{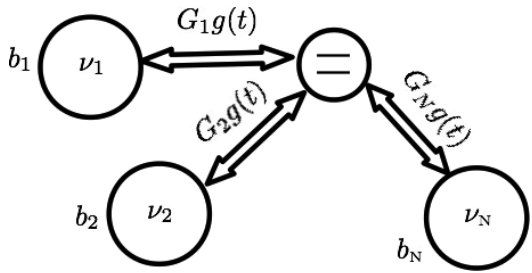}
	\end{center}
	\caption{In the normal modes representation of the oscillator network, the qubit is interacting with those modes $b_k$ such that $G_k\neq0$. Each mode $b_k$ behaves as a simple harmonic oscillator of frequency $\nu_k$, and does not interact with the others [see Eqs.~(\ref{HINT}) and (\ref{H_normal})]. If the assumptions (A1) and (A2) are verified, the qubit interacts with all the normal modes, and can distinguish each mode by its frequency. \label{normalmod}}
\end{figure}

To study the time evolution of the system, it is convenient to first express the network Hamiltonian in the diagonal form 
\begin{equation}
	H_0=\sum_{k=1}^N\nu_kb^\dagger_kb_k,\label{bkbk}
\end{equation}
where $b_k$'s are the normal mode operators with corresponding eigenfrequencies $\nu_k$. We assume that the harmonic network is stable, i.e. $\nu_k>0$ for any $k$. The normal modes $\boldsymbol b=(b_1,...,b_N)$ are related to the local modes $\boldsymbol a=(a_1,...,a_N)$ via \footnote{We use the following conventions in applying conjugations to a column vector of bosonic modes, such as $\boldsymbol a$. The symbol $\boldsymbol a^*$ stands for the column vector where each component is replaced by its adjoint: $\boldsymbol a^*\equiv(a_1^\dagger,...,a_N^\dagger)$, while $\boldsymbol a^\dagger$ stands for the actual adjoint $\boldsymbol a^\dagger\equiv(a_1^\dagger,...,a_N^\dagger)^\intercal$.} 
\begin{align}
	\colvec{\boldsymbol b}{\boldsymbol b^*}=\mathcal S\colvec{\boldsymbol a}{\boldsymbol a^*},\label{b-a}
\end{align}
where $\mathcal S$ is a $2N\times2N$ symplectic matrix \cite{cv_rev}. That is, $\mathcal S$ is a transformation that preserves the canonical commutation relations. It proves convenient to decompose $\mathcal S$ in four $N\times N$ blocks. Looking at \eq{b-a}, we see that it is possible to write down
\begin{equation}
	\mathcal S=\begin{pmatrix}S_1&S_2\\S_2^*&S_1^*\end{pmatrix}\label{s-blocks},
\end{equation}
where $^*$ indicates element-wise complex conjugation (as opposed to hermitian conjugation, where the matrix is also transposed). The preservation of bosonic cummutation relations imply the constraints
\begin{align}
	&S_1^\dagger S_1-(S_2^\dagger S_2)^*=\mathbb1,\label{symplec-1}\\
	&S_1^\dagger S_2=(S_2^\dagger S_1)^*.\label{symplec-2}
\end{align}
As a consequence, the inverse of $\mathcal S$ is given by
\begin{equation}
	\mathcal S^{-1}=\begin{pmatrix}\phantom-S_1^\dagger&-S_2^\intercal\\-S_2^\dagger&\phantom-S_1^\intercal\end{pmatrix}.
\end{equation}
From this, we can express the interaction Hamiltonian of \eq{time-varying} in the new basis:
\begin{align}
	H_\text{int}(t)&=g(t)\sigma_3\sum_{k=1}^N\left(G_kb_k+G_k^*b_k^\dagger \right),\label{HINT}\\
	G_k&=(S_1-S_2)_{k1}^*.\label{G_k}
\end{align}
We note that, in the new representation, the qubit interacts with all the modes $b_k$ such that $G_k\neq0$. Fig.~\ref{normalmod} shows a graphical representation of the Hamiltonian (\ref{HTOT}), in terms of the normal modes of the oscillator network. 
\subsection{Time evolution for a closed system}
It is convenient to evaluate the time evolutor in the normal modes basis. The Hamiltonian (\ref{HTOT}), in an interaction picture with respect to $H_0$, can be recasted as
\begin{align}
H_I(t)&=\sum_{k=1}^N h_k(t),\label{H_normal}\\
h_k(t)&=g(t)\sigma_3\left(G_kb_ke^{-i\nu_kt}+G_k^*b_k^\dagger e^{i\nu_kt}\right).\label{h-small}
\end{align}
We can see that $[h_k(t),h_{k'}(t')]=0$ if $k\neq k'$. Therefore the time evolutor must be of the form $U_I(t)=\otimes_ku_k(t)$, where $u_k$ obeys the Schr\"{o}dinger equation 
\begin{align}
	\dot u_k&=-ih_ku_k,\\
	u(0)&=\mathbb1.
\end{align}
We can impose the ansatz $u_k=e^{i\phi_k}\exp\{\sigma_3[\beta_kb_k^\dagger-\beta_k^*b_k]\}$, where $\phi_k$ and $\beta_k$ are functions of time. This leads to $\dot\beta_k=-ig(t)G_k^*e^{i\nu_k t}$ plus an equation for $\phi_k$ that we do not need to solve, since the product $\prod_k e^{i\phi_k(t)}$ is just a global phase factor that can be ignored. Then, the time evolutor of the system can be given in the closed form:
\begin{equation}
U_I (t)=\exp\left\{\sigma_3[\boldsymbol b^\dagger\boldsymbol\beta(g,t)-\boldsymbol\beta(g,t)^\dagger\boldsymbol b]\right\},\label{evolution}
\end{equation}
where $\boldsymbol\beta(g,t)=(\beta_1(g,t),...,\beta_N(g,t))$ and
\begin{equation}
\beta_k(g,t)=-iG_k^*\int_0^t\mathrm{d}sg(s)e^{i\nu_k s}\label{betak}.
\end{equation}
\section{Realizing arbitrary qubit-controlled displacements}\label{sec_dis}
The time evolutor of Eq. (\ref{evolution}) yields a \textit{qubit-controlled multimode displacement} for the bosonic modes $b_k$, characterized by displacement parameters $\pm\beta_k(g,t)$, the sign being determined by the eigenvalue of $\sigma_3$. Notice that the displacement $\boldsymbol\beta(g,t)$ is a functional of the coupling strength $g(s)$. The ability to tune the latter will be crucial in reconstructing the state of the network. In terms of the local modes $\boldsymbol a$, the time evolutor reads (for brevity, from now on we will omit the explicit dependence of the displacement on $g$ and $t$)
\begin{align}
&U_I(t)=\exp\left\{\sigma_3(\boldsymbol a^\dagger\boldsymbol\alpha-\boldsymbol\alpha^\dagger\boldsymbol a)\right\}\label{locals},
\end{align}
where
\begin{align}
&\colvec{-\boldsymbol\alpha^*}{\boldsymbol\alpha}=\mathcal S^\intercal\colvec{-\boldsymbol\beta^*}{\boldsymbol\beta}.\label{local-normal}
\end{align}
where $\mathcal S^\intercal$ indicates the transpose of $\mathcal S$, while $\boldsymbol\alpha=(\alpha_1,...,\alpha_N)$. At this point we make two assumptions:
\begin{itemize}
	\item (A1) all the coefficients $G_k$ are different from zero
	\item (A2) the normal modes spectrum $\{\nu_1,...,\nu_N\}$ is non-degenerate
\end{itemize}
In physical terms, they imply that the probe qubit \textit{interacts with and can resolve all normal modes} $b_k$ [see Eqs.~(\ref{H_normal},\ref{h-small}) and Fig.~\ref{normalmod}]. 

These assumptions are satisfied by generic networks (\ie, networks without special symmetries) and, in particular, by a linear chain of oscillators. When (A1) and (A2) are verified, it becomes possible to assign arbitrary values to the displacement vector $\boldsymbol\alpha$, just by controlling the length of the interaction time $t$ and by appropriately tailoring the time dependence of the coupling $g(s)$. To see that this is possible, suppose that we wish to apply the operator of Eq. (\ref{locals}), with generic $\boldsymbol\alpha$ of our choice. The corresponding vector $\boldsymbol\beta$ that has to be applied to the normal modes $\boldsymbol b$ is given by 
\begin{equation}
	\colvec{-\boldsymbol\beta^*}{\boldsymbol\beta}=\left(\mathcal S^\intercal\right)^{-1}\colvec{-\boldsymbol\alpha^*}{\boldsymbol\alpha}.
\end{equation} 
Thanks to the assumption (A1), we can impose an interaction strength profile of the form
\begin{equation}
 g(s)=\frac{i}{t}\sum_{l=1}^N\left(\frac{B_l}{G_l^*}e^{-i\nu_l s}-\frac{B_l^*}{G_l}e^{i\nu_l s}\right)\label{analytical},
\end{equation}
with $B_l$ coefficients to be determined. 
Inserting the above expression in Eq.~(\ref{betak}), and defining $\mathbf B\equiv(B_1,...,B_N)$, we obtain
\begin{align}
&\colvec{-\boldsymbol\beta^*}{\boldsymbol\beta}=M\colvec{-\mathbf B^*}{\mathbf B},\label{matrixeq}\\
 &M=\left(\begin{array}{cc}
M_1 & M_2\\
M_3 & M_4
\end{array}\right)\label{m-blocks}
\end{align}
where $M$ is a $2N\times2N$ matrix, whose four $N\times N$ blocks are:
\begin{align}
&(M_1)_{kl}=\frac{G_k}{G_l}\frac{1}{t}\int_0^t\mathrm{d}s e^{-i(\nu_k-\nu_l)s},\label{M1}\\ &(M_2)_{kl}=\frac{G_k}{G^*_l}\frac{1}{t}\int_0^t\mathrm{d}s e^{-i(\nu_k+\nu_l)s},\label{M4}\\
&(M_3)_{kl}=(M_2)_{kl}^*,\\
&(M_4)_{kl}=(M_1)_{kl}^*.
\end{align}
Matrix $M$ is invertible for long enough interaction times. In fact, it is easy to see that assumption (A2) implies
\begin{equation}
\lim_{t\rightarrow\infty}M=\mathbb 1_{2N}\Rightarrow\lim_{t\rightarrow\infty}\det{M}=1\label{limit}
\end{equation} 
where $\mathbb 1_{2N}$ is the $2N\times2N$ identity matrix. This implies that there must be an interaction time $t_0$ such that $\det M>0$ for $t>t_0$. Inverting Eq. (\ref{matrixeq}), we obtain the required values of $(B_1,...,B_N)$:
\begin{equation}
\colvec{-\mathbf B^*}{\mathbf B}=M^{-1}\colvec{-\boldsymbol\beta^*}{\boldsymbol\beta}=\left(\mathcal S^\intercal M\right)^{-1}\colvec{-\boldsymbol\alpha^*}{\boldsymbol\alpha}.\label{B-local}
\end{equation}
A simple Fourier argument as well as the examples we studied numerically suggest that the matrix $M$ is invertible when we take
\begin{equation}
	t>\frac{\pi}{\min_{j\neq k}|\nu_j-\nu_k|},
\end{equation}
that is, when the interaction time is sufficiently long to resolve the smallest frequency difference of the system. In practice, if we require the components of the displacement vector $\boldsymbol\alpha$ (or $\boldsymbol\beta$) to be large, it might be necessary to pick even longer interaction times, since the magnitude $|g(s)|$ is often limited to a maximum value in realistic implementations. In fact, one can see from Eqs.~(\ref{analytical}) and (\ref{B-local}) that, keeping $\boldsymbol\alpha$ (or $\boldsymbol\beta$) fixed, longer interaction times imply a smaller amplitude of the coupling strength $g(s)$. On the other hand, the interaction time $t$ has to be kept small compared to the decoherence timescales of the system. The combination of these two facts imposes practical limits to the maximum attainable values of $|\alpha_n|$ (or $|\beta_k|$), thus reducing the extent of the accessible region in the phase-space of the oscillator network.
\section{Quantum state reconstruction} \label{sec_qsr}
Let us show that the ability to tune the qubit-network coupling allows to reconstruct an arbitrary initial state of the network just by performing measurements on the qubit.
To begin, we initialize the system in the state
\begin{equation}
\rho_\textrm{tot}(0)=\kebra{+}{+}\otimes\rho,\label{initial}
\end{equation}
where $\rho$ is the unknown state of the network at $t=0$ that we want to reconstruct, $\ket +=\tfrac{1}{\sqrt2}(\ket g+\ket e)$ is the positive eigenstate of $\sigma_1$, $\ket e$ and $\ket g$ being respectively the positive and negative eigenstates of $\sigma_3$. We then choose an interaction time $t>t_0$ and a set of local displacement parameters $\boldsymbol\alpha=-\boldsymbol\xi/2$ [with $\boldsymbol\xi=(\xi_1,...,\xi_N)$ and the factor $-\tfrac{1}{2}$ included for later convenience], so that a specific profile of $g(t)$ is determined according to Eqs. (\ref{analytical}) and (\ref{B-local}). After the interaction, the system has evolved to a state $\rho_\textrm{tot}(t)=U_I(t)\rho_\textrm{tot}(0)U_I(t)^\dagger$. We then measure either the qubit observable $\sigma_1$ or $\sigma_2$, and repeat the experiment a sufficient number of times to estimate the average values $\av{\sigma_j}=\tr{\rho_\textrm{tot}(t)\sigma_j}$. By explicit calculation, we get
\begin{align}
&\av{\sigma_1}+i\av{\sigma_2}=\chi(\boldsymbol\xi),\label{chi}
\end{align}
with 
\begin{align}
&\chi(\boldsymbol\xi)\equiv\textrm{tr}\left\{\rho\exp\left(\boldsymbol\xi\cdot\boldsymbol a^\dagger-\boldsymbol\xi^*\cdot\boldsymbol a\right)\right\}\label{chi-def}
\end{align}
being the Weyl characteristic function \cite{barnett} of the oscillator network. By repeating the procedure for different points $\boldsymbol\xi$ of the network phase space, the full characteristic function can be measured. We recall that the latter gives a complete description of the state of a multimode system, equivalent to its Wigner function or density matrix \cite{barnett}. In contrast with standard tomographic reconstructions \cite{qtomo}, Eq.~(\ref{chi}) provides a direct link between $\chi(\boldsymbol\xi)$ and the measured data, without the need of any integral transform of the latter. In a sense, the post-processing typical of quantum tomography is here replaced by the pre-processing needed to determine $g(s)$. The advantage is that, while the former is performed on noisy state-dependent data, the latter involves only the state-independent Hamiltonian parameters. 
As typical for any infinite dimensional system, the full reconstruction of the state $\rho$ [\ie, the entire $\chi(\boldsymbol\xi)$] is impractical. However, a number of interesting properties can be accessed given only a finite collection of $\chi(\boldsymbol\xi)$ values, as we illustrate in the following.

\subsection{Quantum properties without full reconstruction} \label{sec_nofull}
The nonclassicality of a continuous variable state is generally associated with its Wigner function being negative. In turn, a method to probe this nonclassicality criteria directly from a finite collection of characteristic function values has been recently put forward and experimentally tested on a single-mode radiation state \cite{mari}. Our reconstruction scheme is in this respect especially suited, providing directly $\chi(\boldsymbol\xi)$ from measurements. In particular, it might open the way to directly estimate nonclassicality for multi-mode states of massive oscillators. Other nonclassicality criteria, relying on constraints for the P-function, are also testable directly from the characteristic function  \cite{vogel}. In addition, also entanglement can be similarly estimated. In fact, as suggested in Ref.~\cite{mari}, the method outlined there can readily be extended to provide lower bounds for entanglement measures in the multimode setting. 

Dealing with the characteristic function offers other relevant features. For example, it is often the case that one is interested in a block of a system (\ie, its reduced states), rather than the whole system---\eg, to test entropy-area laws for many-body ground states \cite{area}. Given $\chi(\boldsymbol\xi)$, this can be readily done, since tracing away a mode $a_j$ simply corresponds to evaluating $\chi(\boldsymbol\xi|\xi_j=0)$ [see Eq.~(\ref{chi-def})]. More generally, correlation functions are of broad interest---\eg, in many-body model simulations. One can apply polynomial or functional (e.g. Gaussian) fits to a finite set of characteristic function values, measured in the vicinity of the phase-space origin, to estimate low order moments of the modes $\boldsymbol a$ (in dealing with noisy data, this approach is preferable to the extraction of moments by derivatives \cite{barnett}). 
From those, any correlation function of the same order can be calculated. A particular but relevant case appears when one considers Gaussian states. Then, only second moments are necessary to reconstruct the state and properties thereof \cite{cv_rev}. Also, deviation from Gaussianity can be addressed by considering higher order moments. 

\subsection{Temperature measurements}\label{sec_temp}
Let us consider a concrete example of the reconstruction method that might be relevant in first experimental implementations. Suppose that one expects $\rho$ to be in a thermal state of the Hamiltonian of Eq.~(\ref{H_0}), and wants to test this hypothesis. Being thermal states diagonal in the normal modes $\boldsymbol b$, it is convenient to reconstruct the characteristic function directly in terms of the latter---something that can easily be done using the method outlined above. In the normal mode basis, the characteristic function of a generic thermal state is 
\begin{equation}
\bar\chi_T(\boldsymbol\eta)=\exp\left\{-\sum_{k=1}^N\left[\mathcal N(\nu_k)+\frac{1}{2}\right]|\eta_k|^2\right\},\end{equation}
where $\mathcal N(\nu_k)=1/(e^{\nu_k/T}-1)$
is the number of bosonic excitations at frequency $\nu_k$ and temperature $T$. Following the procedure above, one can choose a set of displacement parameters $\boldsymbol\beta=-\boldsymbol\eta/2$ [see Eq.~(\ref{B-local})], so that a finite collection of $\chi(\boldsymbol\eta)$ values can be reconstructed directly in the normal mode basis. Then, standard statistical methods can be employed both to test the validity of the thermal hypothesis and estimate $T$.

\section{Noise and errors}\label{sec_noise}
Let us discuss some of the main sources of error that could affect our scheme. Firstly, there is the unavoidable coupling of the system to the external environment, giving rise to decoherence. If this effect can be modelled via a standard Markovian master equation, the state of the network can still be reconstructed in full detail, at the expense of collecting larger amounts of statistical data. Secondly, systematic errors might limit the precision to which we can control the coupling $g(t)$, meaning that the actual displacement parameters will be slightly different from the desired values. This will effectively limit the phase-space resolution of the reconstructed state. These two important sources of error are discussed in detail in the sections below.

Another source of error arises from the experimental uncertainties in the Hamiltonian parameters in Eq.~(\ref{H_0}), and it affects every stage of our protocol through standard error propagation. It is then crucial for the assumptions (A1) and (A2) to be verified for the whole range of parameters inside the error bars. If this condition is met, the inversion of matrix $M$ in Eq.~(\ref{B-local}) remains well defined. Moreover, Eq.~\eqref{limit} guarantees that, for long enough interaction times, the uncertainty on $M^{-1}$ will be of the same order of the uncertainty on $M$.

Finally, as common in many reconstruction protocols, errors in the measured data can yield a non-physical reconstructed state. In our case, the crucial issue is to check whether a finite collection of measured characteristic function values (with associated uncertainties) is compatible with a positive semidefinite density matrix [normalization can be satisfied simply by imposing $\chi(0)=1$].
This problem can be addressed directly by making use of the quantum Bochner Theorem \cite{quantum-bochner}, for example by using the numerical methods developed in Ref.~\cite{mari}. There it is shown how, by using semidefinite programming, it is possible to output a set of characteristic function values, compatible with the data but devoid of non-physicalities.  
\subsection{Markovian Decoherence}
To treat environmental noise in our model, it is convenient to work in the normal modes basis of the oscillator network. To simplify matters, we will restrict the discussion to the decoherence of the oscillators being ``diagonal" in terms the normal modes $b_k$. This can be a good model of decoherence when the environmental noise is completely uncorrelated between different nodes of the network, and when the inter-oscillator couplings $K_{nm}$ are small with respect to the local frequencies $\omega_n,\omega_m$ (see Appendix \ref{appendix1} for a more detailed discussion)
A widely applicable model of Markovian decoherence for both the qubit and the oscillators is given by the master equation \cite{barnett}
\begin{align}
&\dot\rho_\text{tot}=-i[H_I(t),\rho_\text{tot}]+\sum_k L_k\rho_\text{tot}+\mathcal Q\rho_\text{tot}.\label{master-multi}
\end{align}
The terms responsible for decoherence are
\begin{align}
&L_k=\frac{\kappa_k}{2}(\mathcal N_k+1)\mathcal D[b_k]+\frac{\kappa_k}{2}\mathcal N_k\mathcal D[b_k^\dagger],\label{lind-multi}\\
&\mathcal Q=\frac{\Gamma_1}{2}(\mathcal N_q+1)\mathcal D[\sigma_-]+\frac{\Gamma_1}{2}\mathcal N_q\mathcal D[\sigma_+]+\frac{\Gamma_2}{2}\mathcal D[\sigma_3],\label{lind-3}
\end{align}
where $\sigma_+=\kebra{e}{g},\sigma_-=\kebra{g}{e}$, $\kappa_k$ $(k=1,...,N)$ is the coupling of each normal mode to the environment (they might be in general different, as each normal mode has a different frequency), while $\mathcal N_k=\mathcal N(\nu_k)$ is the thermal occupation of the environment at frequency $\nu_k$. $\Gamma_1$ and $\Gamma_2$ are the qubit couplings to the environment, the first being responsible for thermalization, $\mathcal N_q=\mathcal N(\omega_q)$ being the thermal occupation of the environment at frequency $\omega_q$, while $\Gamma_2$ is the strength of additional dephasing mechanisms. 
Finally, the action of the superoperator $\mathcal D$ on a generic operator $A$ is
\begin{equation}
\mathcal D[A]\rho_\text{tot}=2 A\rho_\text{tot} A^\dagger- A^\dagger A\rho_\text{tot}-\rho_\text{tot}A^\dagger A.
\end{equation}
Note that, by using the above model of decoherence, we have the simplification that the environment does not induce any coupling between the normal modes of the network. (see Appendix \ref{appendix1}).

By solving the dynamics analytically, it can be shown (see Appendix \ref{appendix2}) that \eq{chi} has to be modified as follows:
\begin{equation}
	\av{\sigma_1}+i\av{\sigma_2}=\chi(\boldsymbol\eta)e^{-f(g,t)},\label{chi-deco}
\end{equation}
$f$ being a positive function:
\begin{equation}
	f(g,t)=\gamma t+\sum_k\left[\Delta_k\left(1-e^{-\kappa_k t}\right)|\mu_k(t)|^2+\tau_k(t)\right],
\end{equation}
where the explicit forms for $\mu_k$ and $\tau_k$ are given by Eqs.~\eqref{mu-n_t} and \eqref{tau-n_t} respectively.
We can see that the effect of decoherence is twofold. Firstly the matrix $M$, which in this case gives $(-\boldsymbol\eta^*,\boldsymbol\eta)=-2M(-\mathbf B^*,\mathbf B)$, is now given by 
\begin{align}
	&(M_1)_{kl}=\frac{G_k}{G_l}\frac{1}{t}\int_0^t ds e^{-i(\nu_k-\nu_l)s-\frac{\kappa_k}{2}s},\label{M1-deco}\\
	&(M_2)_{jk}=\frac{G_k}{G^*_l}\frac{1}{t}\int_0^t ds e^{-i(\nu_k+\nu_l)s-\frac{\kappa_k}{2}s}\;,
	\label{M2-deco}
\end{align}
while $(M_3)_{kl}=(M_2)_{kl}^*$, $(M_4)_{kl}=(M_1)_{kl}^*$. A crucial point in our protocol is the invertibility of the matrix $M$, since it allows us to assign arbitrary values to $\boldsymbol\eta$. We have already seen that if the time $t$ is chosen large enough, this matrix can be inverted in the case $\kappa_k=0$, \ie,  det$M\neq0$. Due to the continuity of the determinant, if the condition $t\ll1/\kappa_k$, $(k=1,...,N)$ can be verified, then $M$ is just slightly perturbed when $\kappa_k\neq0$, so that its determinant remains different from zero. In cases where $t$ and $1/\kappa_j$ are of the same order, the invertibility of the matrix $M$ should be verified numerically. The second effect of decoherence is the appearance of the damping term $e^{-f}$ in Eq.~(\ref{chi-deco}), meaning that the measured quantity deviates from the actual value of the characteristic function. However, since the function $f(g,t)$ is state-independent in our model, and $0\leq f<\infty$, it follows that the right hand side of Eq.~(\ref{chi-deco}) is still a valid representation of the quantum state $\rho$. This means that we could in principle recover the value of $\chi(\boldsymbol\eta)$, if the decoherence parameters of the system are known with sufficient accuracy, simply by multiplying the measured data by $e^f$. Note however that this operation also applies to the experimental uncertainty, so that an error $\delta$ in the measured data implies a larger error $\delta e^{f}$ in the knowledge of $\chi(\boldsymbol\eta)$. To compensate for this, the expectation values of the Pauli operators in Eq.~(\ref{chi-deco}) have to be measured with higher accuracy compared to the decoherence-free case, which necessarily requires a larger number of experimental repetitions. Finally, we recall that we are considering $\chi$ in the normal modes basis, so the complex vector $\boldsymbol\eta$ is related to the local modes vector $\boldsymbol\xi$ via $(-\boldsymbol\xi^*,\boldsymbol\xi)=\mathcal S^\intercal(-\boldsymbol\eta^*,\boldsymbol\eta)$.
\subsection{Noise in the tunable coupling}
Since our protocol relies heavily on the controllability of the time-dependent coupling $g(t)$, it is natural to question its robustness against imprecisions in such control. We model systematic errors in the controllable coupling by substituting 
\begin{equation}
g(s)\rightarrow g(s)+\zeta(s),\label{gnoise}
\end{equation}
where $\zeta$ is a small white noise component:
\begin{align}
	&\overline{\zeta(s)}=0,\label{zetamean}\\
	&\overline{\zeta(t_1)\zeta(t_2)}=\epsilon\delta(t_1-t_2),\label{zetavar}
\end{align}
and $\overline{\phantom{00}}$ indicates the ensemble average, that is, the average over many realizations of the noise [physically, over many repetitions of the experiment, where the deterministic component $g(s)$ is kept fixed]. In \eq{zetavar}, $\epsilon$ has the dimensions of a coupling strength variance per unit of frequency (with our choice of units, this amounts to the dimensions of a frequency), and it represents the strength of the white noise. Loosely speaking, $\epsilon$ represents the ``thickness" of the curve $(s,g(s))$.
As a result of \eq{gnoise}, at each repetition of the experiment the displacement parameters in the time evolutor of Eqs.~(\ref{evolution}) and (\ref{locals}) deviate from the desired values by a small random amount. This yields a finite resolution in our power of observation of the oscillator phase-space, meaning that we will only be able to measure a coarse-grained version of the Characteristic Function, where sub-resolution features are washed out. For simplicity, let us compute this phase-space resolution in the normal modes basis. If we plug \eq{gnoise} into \eq{betak}, we see that the displacement parameters $\beta_k$ in the time evolutor (\ref{evolution}) are replaced by
\begin{equation}
	\beta_k\rightarrow\beta_k+\delta\beta_k,
\end{equation}
where $\delta\beta_k$ are the complex random variables
\begin{equation}
	\delta\beta_k=-iG_k^*\int_0^t\mathrm{d}s\zeta(s)e^{i\nu_k s}.
\end{equation}
Due to Eqs.~(\ref{zetamean}) and (\ref{zetavar}), we can see that their means and covariances are respectively
\begin{align}
	\overline{\delta\beta_k}&=0,\\
	[V(\delta\boldsymbol\beta)]_{kk'}&=\frac{1}{2}\left(\overline{\delta\beta_k\delta\beta_{k'}^*+\delta\beta_k^*\delta\beta_{k'}}\right)\nonumber\\
	\phantom{[V(\delta\boldsymbol\beta)]_{kk'}}&=\epsilon\phantom0\operatorname{Re}\left\{G_k^*G_{k'}\int_0^t\mathrm{d}se^{i(\nu_k-\nu_{k'})s}\right\}.\label{dbetavar}
\end{align}
By diagonalizing $V(\delta\boldsymbol\beta)$, we can find a new basis for the phase-space, such that the noise along different (orthogonal) directions is uncorrelated. Note how such diagonalization is independent on the noise strength $\epsilon$, and it only depends on the interaction time $t$ and the structure of the network. The square roots of the eigenvalues of $V(\delta\boldsymbol\beta)$ can then be used as a measure of the achievable phase-space resolution, along the phase-space directions defined by the new basis. When the interaction time is large enough, one can see that the diagonal terms $[V(\delta\boldsymbol\beta)]_{kk}$ are dominant\footnote{This happens when considering an interaction time $t\gg1/\text{min}_{j\neq k}|\nu_j-\nu_k|$, similarly to what we had for the invertibility of the matrix $M$.}. Thus, to have an estimate of the order of magnitude of our accuracy in phase-space, we can look at the diagonal elements of the covariance matrix, which have the simple form:
\begin{equation}
	[V(\delta\boldsymbol\beta)]_{kk}=|G_k|^2\epsilon t.
\end{equation}
It follows that, as a first approximation, and assuming that the presence of systematic noise in the coupling strength is well modelled by \eq{gnoise}, the smallest resolvable feature in the phase space of the oscillator network scales only sublinearly with $\epsilon$ and $t$: 
\begin{equation}
	|\delta\beta_k|\sim|G_k|\sqrt{\epsilon t}.
\end{equation}
\section{Example: Linear chain with constant couplings}\label{linearchain}
\begin{figure}
\begin{center}\includegraphics[width=.48\linewidth]{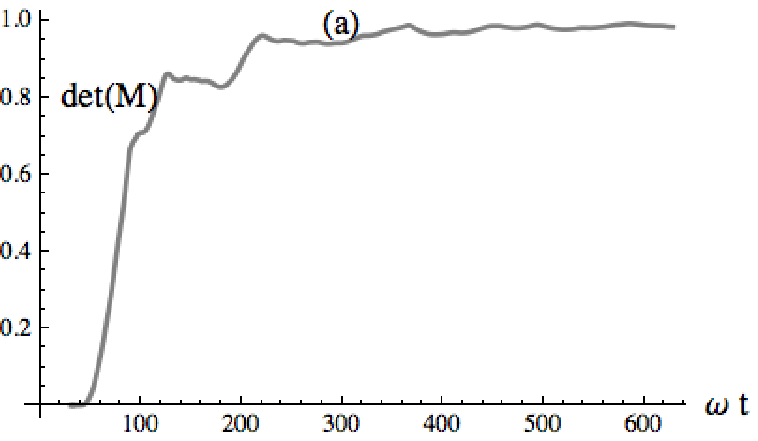}\hspace{.04\linewidth}\includegraphics[width=.48\linewidth]{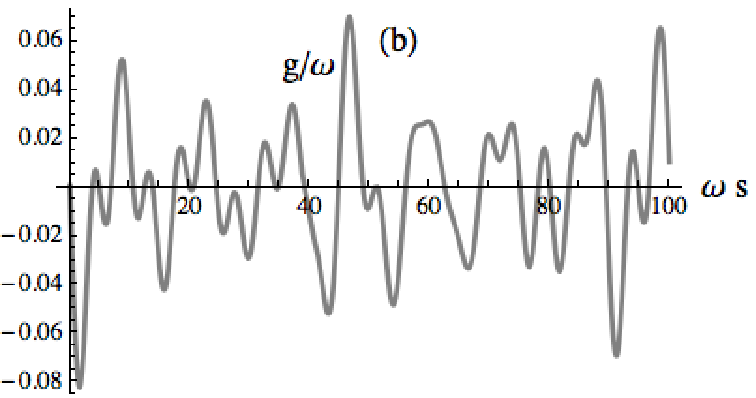}
		\vspace{.1cm}
	\includegraphics[width=.48\linewidth]{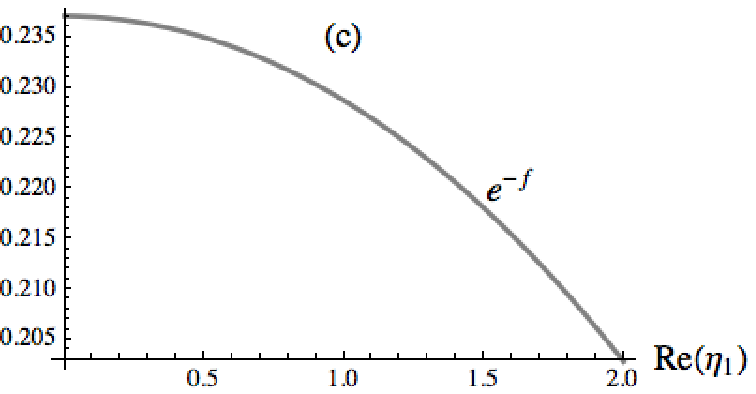}\includegraphics[width=.48\linewidth]{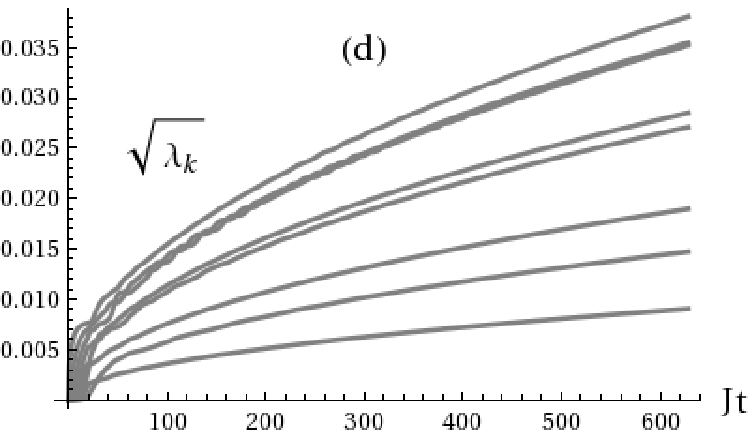}
	\end{center}
\caption{\label{chainplots} Reconstruction protocol for a linear chain of $N=8$ oscillators. In all plots we consider nearest-neighbour couplings $J_{nm}=K_{nm}=J\delta_{m,n+1}$, with $J=0.2\omega$, decoherence rates $\kappa_k=\kappa=10^{-6}\omega$ temperature $T=200\omega$ and a coupling noise strength $\epsilon=10^{-5}\omega$. Note that we are requiring a motional quality factor $Q\sim\tfrac{\omega}{\kappa\mathcal N}\sim5\times10^{3}$. We assumed the decoherence of the qubit to be negligible. Plot (a) shows the determinant of the matrix $M$, which we require to be different from zero in our protocol. We see that $M$ becomes invertible for $t\gtrsim50/\omega$. In plots (b) and (c), we have considered a phase-space region $|\eta_k|\leq2$ in the normal modes basis, and the total interaction time has been fixed to $t=100\times2\pi/\omega$. Plot (b) shows a portion of the interaction strength profile required to obtain the displacement parameters $\boldsymbol\beta=(-1,-1,...,-1)$, which allows us to reconstruct the phase space point $\boldsymbol\eta=(2,2,...,2)$, lying at the boundary of the considered region. With our choice of parameters, it is necessary to access maximal coupling strengths of ${|g|_{\text{max}}}\simeq0.08\omega$ to reach this phase space point. In plot (c), the effect of decoherence on the measured value of the characteristic function is shown. We fixed $\eta_2=\eta_3=...=\eta_N=2$ and $\operatorname{Im}\{\eta_1\}=0$. The plot shows the quantity $e^{-f}$ as a function of $\operatorname{Re}\{\eta_1\}$. We see that near the boundary of the considered phase-space region the quantity measured via the qubit expectation values corresponds to about 20-23$\%$ of the actual value of the characteristic function [see \eq{chi-deco}]. Plot (d) shows the maximum achievable phase-space resolution along the directions that diagonalize $V(\delta\boldsymbol\beta)$ [we take the square roots of its eigenvalues $\lambda_k$ as a measure of the phase-space accuracy]. The asymptotical behaviour for large $t$ is given by $\sqrt{\lambda_k}\simeq|G_k|\sqrt{\epsilon t}$, as expected. As a final remark, we note that the considered range of interaction times is consistent with Eq.~(\ref{master-condition}), which gives a necessary condition for the validity of the master equation used. Indeed, from Eq.~(\ref{master-condition}) one can estimate that our treatment is valid for  $t\ll t_\textrm{max}$, where $t_\textrm{max}
=\min_{k}\left[\frac{1}{\kappa_k\mathcal N_k}\right]\frac{\omega}{K}\sim2\times10^3(2\pi/\omega)$.}
\end{figure}

Let us give a concrete example of a quantum oscillator network where the presented ideas can be applied. Consider a linear chain of $N$ oscillators, each having the same local frequency $\omega$, and where only the nearest-neighbours interact with a coupling strength $J_{n,n+1}=K_{n,n+1}=J$, constant along the chain. The $J$'s are often referred to as \textit{hoppings}. We assume that the qubit is tunably coupled to the first oscillator of the chain. The system is sketched in Fig.~\ref{chain}:
\begin{figure}
	\begin{center}
		\includegraphics[width=.8\linewidth]{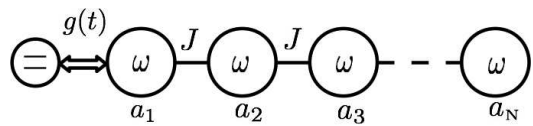}
	\end{center}
\caption{Sketch of a qubit tunably coupled to a linear chain of oscillators with constant nearest-neighbour couplings.\label{chain}}
\end{figure}
In terms of the parameters appearing in the Hamiltonian $H_0$ [see \eq{H_0}], we have
\begin{align}
	\omega_n&=\omega\\
	K_{nm}=J_{nm}&=\left\{\begin{array}{lr}
	J\delta_{m,n+1}&\qquad n<N,\\
	0&\qquad\text{otherwise}.
	\end{array}\right.
\end{align}
The diagonalization of the Hamiltonian $H_0$ can be performed analytically, yielding the spectrum (here $k=1,...,N$)
\begin{align}
	&\nu_k=\sqrt{\omega(\omega+2\varepsilon_k)},\label{chain-spectrum}\\
	&\varepsilon_k=2J\cos\left(\frac{\pi k}{N+1}\right).\label{epsilon-def}
\end{align}
One can see that the above spectrum is non-degenerate, thanks to the fact that the cosine is monotone in the interval $[0,\pi]$. Thus, the linear chain of oscillators verifies the assumption (A2). The symplectic matrix $\mathcal S$, connecting the local modes $a_n$ to the normal modes $b_k$, can also be expressed in analytical form. Its main blocks $S_1$ and $S_2$ are (see Section~\ref{normal-modes})
\begin{align}
	&(S_1)_{kn}=\sqrt{\frac{2}{N+1}}\cosh{(r_k)}\sin\left(\frac{\pi k n}{N+1}\right),\label{S_1chain}\\
	&(S_2)_{kn}=-\sqrt{\frac{2}{N+1}}\sinh{(r_k)}\sin\left(\frac{\pi k n}{N+1}\right),\label{S_2chain}\\
	&r_k=\tanh^{-1}\left(\frac{\varepsilon_k}{\omega+\varepsilon_k+\nu_k}\right).
\end{align}
If we combine Eqs.~(\ref{S_1chain}), (\ref{S_2chain}) and (\ref{G_k}), we have
\begin{equation}
	G_k=\sqrt{\frac{2}{N+1}}e^{r_k}\sin\left(\frac{\pi k}{N+1}\right),
\end{equation}
which is different from zero for any $k\in(1,...,N)$. Therefore, also assumption (A1) is verified. Thus, the quantum state of a linear chain of oscillators with constant nearest-neighbour couplings can be fully reconstructed, by using a single qubit coupled to one end of the chain. 
As an example, Fig.~\ref{chainplots} shows some quantities of interest for the reconstruction protocol of a linear chain of $N=8$ oscillators.
\section{Possible experimental implementations}\label{sec_exp}
The reconstruction scheme described here is based on a rather ubiquitous dynamics. Essentially, it requires a harmonic coupling between the oscillators and a bilinear qubit-oscillator interaction. An experimental platform that is particularly mature for our purposes is given by a chain of ions in a linear trap. There the harmonic dynamic is provided by the Coulomb interaction, as recently demonstrated in Ref.~\cite{2ions}. In addition, the required qubit-oscillator coupling is standard \cite{ion_rmp}: two electronic levels of one ion provide the qubit, whereas the coupling is realized via a standing laser wave \cite{james}. For example, consider a linear chain of $N=8$ ions, with $\omega_n\sim\omega, J_{n,m}=K_{nm}\sim\delta_{m,n+1}0.2\omega$ and take an interaction time $t\sim100(2\pi/\omega)$. Then $|g(s)|_{\max}\sim0.08\omega$ allows to reconstruct states with $\chi(\boldsymbol\xi)$ having support in $|\xi_n|\lesssim2$. Notice that a modest motional quality factor $Q\gtrsim5\times10^3$ is required (see Fig.~\ref{chainplots}), and it is sufficient to vary the laser power on a time scale of the eigenfrequencies $\nu_k$ (typically of the order of MHz) in order to realize the desired profile of $g(s)$. We stress again that only one ion needs to be illuminated in order to reconstruct the motional state of the entire chain. 
\par
Similar couplings can be envisaged also in a circuit QED setup \cite{Blais}, where many stripline waveguides can be capacitively coupled mimicking a harmonic network, and a single superconducting qubit can be coupled to one of the resonators. In this setting, the realization of the Hamiltonian (\ref{H_normal}) requires a regime where the qubit level splitting is negligible compared to the dipolar coupling. Ref.~\cite{cirQEDtuning} is promising in this direction, as it demonstrates independent tunability of these two parameters.

Being our model quite general, its implementation can be envisaged also in other experimental platforms, such as nanomechanical oscillators or microcavities. In fact, the recent effort to build complex quantum networks led to impressive experimental progresses. In this context, the question of extracting information from a quantum network by only accessing a limited portion of it is of interest from a general viewpoint. We provided here a solution to this problem for a network of quantum oscillators and we expect that further investigations for the case of different constituents will be relevant in the future.

\textit{Acknowledgments} --- We thank D. Burgarth, W. Hensinger, A. Mari, A. Retzker and X. Wang for the useful discussions. We acknowledge support from the UK EPSRC, the QIPIRC, the EU under Marie Curie IEF No 255624, the Royal Society and the Wolfson Foundation.
\appendix
\section{Derivation of the master equation for the oscillator network}\label{appendix1}
Let us start by writing down a Hamiltonian that includes the harmonic oscillator network, its environment, and their mutual interaction. We assume that each oscillator mode $a_n$ is interacting with a local environment, described by a continuum of bosonic modes $c_n(\omega)$, and that their interaction is bilinear, with frequency-dependent coupling strength $f_n(\omega)$. Thus we take
\begin{align}
	&H_\text{tot}=H_0+H_{env}+H_{SE},\label{htot-env}\\
	&H_0=\sum_{n=1}^N\omega_na^\dagger_na_n+\sum_{n<m}J_{nm}\left(a_na_m^\dagger+a_n^\dagger a_m\right)+\nonumber\\
	&\phantom{H_0}+\sum_{n<m}K_{nm}\left(a_na_m+a_n^\dagger a_m^\dagger\right),\label{H_0A}\\
	&H_{SE}=\sum_{n}(a_n+a_n^\dagger)\int \mathrm{d}\omega f_n(\omega)[c_n(\omega)+c_n^\dagger(\omega)],\label{env-coupling}\\
	&H_{env}=\sum_{n}\int \mathrm{d}\omega \phantom{0}\omega c_n^\dagger(\omega)c_n(\omega),
\end{align}
We now assume that the environment is completely uncorrelated between different points of the oscillator network, and that the corresponding bosonic modes are independent, that is:
\begin{align}
	&[c_n(\omega),c_m^\dagger(\omega')]=\delta_{nm}\delta(\omega-\omega').
\end{align}
If in \eq{htot-env} we switch to the normal modes $b_k$, which diagonalize $H_0$, we can rewrite the interaction Hamiltonian of \eq{env-coupling} as
\begin{align}
	&H_{SE}=\sum_{kn}(\vartheta_{nk}b_k+\vartheta_{nk}^*b_k^\dagger)\int \mathrm{d}\omega f_n(\omega)[c_n(\omega)+c_n^\dagger(\omega)],\\
	&\vartheta=(S_1-S_2)^*,\label{thetadef}
\end{align}
where $S_1,S_2$ are the $N\times N$ matrices defined in \eq{s-blocks}. Assuming that the coupling between system and environment is weak, we can neglect the counter-rotating terms \cite{james-effective}
\begin{equation}
	H_{SE}\simeq\sum_{kn}\int \mathrm{d}\omega f_n(\omega)[\vartheta_{nk}^*b_k^\dagger c_n(\omega)+\vartheta_{nk}b_kc_n^\dagger(\omega)].
\end{equation}
We take the further approximation that the environmental coupling is the same at any location of the network: $f_n(\omega)=f(\omega)$, so that we can perform the sum over the index $n$, and write
\begin{align}
	&H_{SE}\simeq\sum_k\int \mathrm{d}\omega f(\omega)[b_kd_k^\dagger(\omega)+b_k^\dagger d_k(\omega)],\\
	&d_k(\omega)=\sum_n\vartheta_{nk}^*c_n(\omega).
\end{align}
Note that the operators $d_k$ are not bosonic, since the matrix $\vartheta$ is not unitary in general. Indeed, \eq{thetadef} together with Eqs.~(\ref{symplec-1}) and (\ref{symplec-2}) imply
\begin{equation}
	\vartheta^\dagger \vartheta=\mathbb1-(S_2^\dagger S_1)^*-(S_1^\dagger S_2)^*+2(S_2^\dagger S_2)^*.
\end{equation}
However, if the ``active" terms in the Hamiltonian are weak, that is $|K_{nm}|\ll\omega_{n},\omega_m$ [for any combination of $n,m$ --- see \eq{H_0A}], then one can see that 
\begin{equation}
	S_2\sim\mathcal O\left(\frac{K}{\omega}\right),
\end{equation} 
where the expression $\mathcal O\left(\frac{K}{\omega}\right)$ indicates the order of magnitude of the ratios $K_{nm}/\omega_n,K_{nm}/\omega_m$. It follows that $\vartheta$ is approximately unitary and the operators $d_k$ become approximately bosonic. \footnote{This is a consequence of the fact that, in the considered approximation, the interaction terms proportional to $K_{nm}$ in \eq{H_0A} are fast-rotating and can be neglected. Consequently, to a first approximation the symplectic transformation $\mathcal S$, connecting the modes $a_n$ to the modes $b_k$, does not mix creation and annihilation operators with each other, hence $S_2\sim0$. The first correction to this is at least first order in the ratios $K_{nm}/(\omega_n,\omega_m)$, so that $S_2\sim\mathcal O(K_{nm}/\omega_p)$.}
\begin{align}
	&\vartheta^\dagger \vartheta=\mathbb 1+\mathcal O\left(\frac{K}{\omega}\right),\\
	&[d_k(\omega),d_{k'}^\dagger(\omega')]=\delta_{kk'}\delta(\omega-\omega')+\mathcal O\left(\frac{K}{\omega}\right).
\end{align}
Then, the free Hamiltonian of the environment can be rewritten as
\begin{equation}
	H_\text{env}=\sum_k\int \mathrm{d}\omega\phantom0\omega d^\dagger_k(\omega)d_k(\omega)+\mathcal O\left(\frac{K}{\omega}\right).
\end{equation}
It follows that, at the zeroth order in $K_{nm}/\omega_q$, each $k$-subspace evolves independently, according to a total Hamiltonian
\begin{align}
&H_k\simeq \nu_kb^\dagger_k b_k+\int\mathrm{d} \omega f(\omega)[b_kd_k^\dagger(\omega)+b_k^\dagger d_k(\omega)]+\nonumber\\
	&\phantom{H_k}+\int\mathrm{d} \omega\phantom0\omega d_k^\dagger(\omega)d_k(\omega),
\end{align}
where the operators $d_k$ can be treated as bosonic. At this point, one can apply standard techniques to derive separately the master equation for each normal mode $b_k$ \cite{barnett}. Putting together all the modes, one can easily derive the oscillator part of the master equation of \eq{master-multi}. In particular, the coupling parameters $\kappa_k$ are given by
\begin{equation}
	\kappa_k=2\pi [f(\nu_k)]^2.
\end{equation}
Due to the approximations used to derive it, our master equation is only valid for timescales such that
\begin{equation}
	t\ll\min_{k}\left[\frac{1}{\kappa_k\mathcal N_k}\right]\mathcal O\left(\frac{\omega}{K}\right),\label{master-condition}
\end{equation}
where $\mathcal N_k$ is the number of thermal bosonic excitations at frequency $\nu_k$.
\section{Solving the master equation}\label{appendix2}
To solve the master equation, we consider a representation in which a matrix of characteristic functions is used to describe the state of the coupled system. We decompose the total density matrix at time $t$ as
\begin{align}
&\rho_\textrm{tot}(t)=\rho_e(t)\otimes\kebra{e}{e}+\rho_g(t)\otimes\kebra{g}{g}+\nonumber\\
&\phantom{\rho_\textrm{tot}(t)}+\rho_+(t)\otimes\kebra{e}{g}+\rho_-(t)\otimes\kebra{g}{e},
\end{align}
where the operators $\rho_j$ $(j=e,g,+,-)$ belong to the oscillators. If we define the characteristic function for each element as
\begin{equation}
\chi_j(\boldsymbol\beta,t)=\text{tr}_{b_1,...,b_N}\left\{\rho_j(t)\exp(\boldsymbol{b}^\dagger\boldsymbol\beta-\boldsymbol\beta^\dagger\boldsymbol b)\right\},
\end{equation}
we can write
\begin{align}
&\boldsymbol\chi(\boldsymbol\beta,t)=\chi_e(\boldsymbol\beta,t)\kebra{e}{e}+\chi_g(\boldsymbol\beta,t)\kebra{g}{g}+\nonumber\\
&\phantom{\boldsymbol\chi(\boldsymbol\beta,t)}+\chi_+(\boldsymbol\beta,t)\kebra{e}{g}+\chi_-(\boldsymbol\beta,t)\kebra{g}{e}.
\end{align}
In this formalism, the expectation values of the Pauli operators in Eq.~(15) of the main text yield:
\begin{align}
&\text{tr}\{\rho_\text{tot}(t)(\sigma_1+i\sigma_2)\}=2\chi_-(0,t)
\end{align}
It follows that we only need to compute the time evolution of the element $\chi_-$. By using standard techniques \cite{barnett}, we can convert the bosonic operators $b_k$ and $b_k^\dagger$ into differential operators for the characteristic function. Applying this procedure to our master equation we find that $\chi_-$ is decoupled from the other elements, and obeys the following equation:
\begin{align}
	&\partial_t\chi_-=-2i\sum_kg(t)(G_ke^{-i\nu_k t}\partial_{\beta^*_k}-G_k^*e^{i\nu_k t}\partial_{\beta_k})\chi_-+\nonumber\\
	&\phantom{\partial_t\chi_-}+\sum_k\mathcal L_k\chi_--\gamma\chi_-,\label{chi+multi}\\
	&\mathcal L_k=-\frac{\kappa_k}{2}\left(\beta_k\partial_{\beta_k}+{\beta^*_k}\partial_{\beta^*_k}+2\Delta_k|\beta_k|^2\right),\\
	&\Delta_k=\mathcal N_k+\frac{1}{2},\quad\gamma=\Gamma_1(\mathcal N_q+1/2)+2\Gamma_2.
\end{align}
The corresponding solution is
\begin{align}
&\chi_-(\boldsymbol\beta,t)=\chi_-\left(\tilde{\boldsymbol\beta}(t)+\boldsymbol\eta(t),0\right)e^{-\gamma t}\times\nonumber\\
&\phantom{\chi_-(\boldsymbol\beta,t)}\times\prod_ke^{-\Delta_k\left(1-e^{-\kappa_k t}\right)|\beta_k+\mu_k(t)|^2-\tau_k(t)},\label{chi+multi-sol}\\
&\tilde{\boldsymbol\beta}(t)=\left(\beta_1e^{-\frac{\kappa_1}{2}t},...,\beta_Ne^{-\frac{\kappa_N}{2}t}\right),\\
&\boldsymbol\eta(t)=\left(\eta_1(t),...,\eta_N(t)\right),\\
&\eta_k(t)=2iG_k^*\int_0^tdsg(s)e^{i\nu_k s-\frac{\kappa_k}{2}s}\label{xi-n_t},\\
&\mu_k(t)=\frac{2iG_k^*}{\sinh{\frac{\kappa_k}{2}t}}\int_0^tdsg(s)e^{i\nu_k s}\sinh{\frac{\kappa_k}{2}s}\label{mu-n_t},\\
&\tau_k(t)=\kappa_k\Delta_k\int_0^tds|\mu_k(s)|^2\label{tau-n_t}.
\end{align}
The initial state of Eq.~\eqref{initial} of the main text implies the initial condition $\chi_-(\boldsymbol\beta,0)=1/2\chi(\boldsymbol\beta)$, where $\chi(\boldsymbol\beta)$ is the characteristic function of the initial state of the oscillators, expressed in the normal modes basis. Then, \eq{chi-deco} follows by substituting $\boldsymbol\beta=0$ in \eq{chi+multi-sol}


\begin{thebibliography}{99}
\bibitem{note1} By confined quantum modes we mean that each oscillator of the network is confined. However, the network can sustain travelling modes, as e.g. phonons in a crystal lattice of confined atoms.
\end{thebibliography}
\end{document}